\documentclass[aps,prl,groupedaddress,twocolumn,showpacs]{revtex4-1}
\usepackage{graphicx}% Include figure files
\usepackage{bm}

\begin{document}

% Use the \preprint command to place your local institutional report
% number in the upper righthand corner of the title page in preprint mode.
% Multiple \preprint commands are allowed.
% Use the 'preprintnumbers' class option to override journal defaults
% to display numbers if necessary
%\preprint{}

%Title of paper
\title{Absence of Spin Liquid in the Hubbard model on the Honeycomb Lattice}
%\title{A study on the possible emergence of a spin liquid in unfrustrated correlated systems by quantum cluster method}

% repeat the \author .. \affiliation  etc. as needed
% \email, \thanks, \homepage, \altaffiliation all apply to the current
% author. Explanatory text should go in the []'s, actual e-mail
% address or url should go in the {}'s for \email and \homepage.
% Please use the appropriate macro foreach each type of information

% \affiliation command applies to all authors since the last
% \affiliation command. The \affiliation command should follow the
% other information
% \affiliation can be followed by \email, \homepage, \thanks as well.
\author{Atsushi Yamada}
\email[]{atsushi@faculty.chiba-u.jp}
%\homepage[]{Your web page}
%\thanks{}
%\altaffiliation{}
\affiliation{Department of Physics, Chiba University, Chiba 263-8522, Japan}
%Collaboration name if desired (requires use of superscriptaddress
%option in \documentclass). \noaffiliation is required (may also be
%used with the \author command).
%\collaboration can be followed by \email, \homepage, \thanks as well.
%\collaboration{}
%\noaffiliation

\date{\today}

%
% extractbb fig1v3.pdf
%
%

\begin{abstract}
The possible emergence of a spin liquid phase in the half-filled Hubbard model on the honeycomb lattice; a simple model of graphene, is studied 
using the variational cluster approximation. 
We found that the critical interaction strength of a magnetic transition is slightly lower than that of the non-magnetic metal-to-insulator 
transition and the magnetic order parameter is already non-negligible at the latter transition point. 
Thus a semi-metallic state becomes a magnetic insulator as the interaction strength increases 
and a spin liquid state, characterized by an insulating state without long range order, 
or a state very close to that is not realized in this system.   
\end{abstract}

% insert suggested PACS numbers in braces on next line

% Metal-insulator transition, 
% Lattice fermion models (Hubbard model, etc.)
% Strongly correlated electron systems; heavy fermions
% Theories and models of many-electron systems

% insert suggested keywords - APS authors don't need to do this
%\keywords{}

%\maketitle must follow title, authors, abstract, \pacs, and \keywords
\maketitle

% body of paper here - Use proper section commands
% References should be done using the \cite, \ref, and \label commands
{\it Introduction.}--- 
A spin liquid state, which is insulating without showing any long range order, has attracted a lot of interests. 
An insulator without long range order is fairly common among band insulators, 
where the effect of the electron correlations are very small. An important difference of a spin liquid state from band insulators is that 
a spin liquid state is insulating due to the effect of electron correlations. 
In many cases, the electron correlations derive a system to a state with long range order, e.g., a magnetic state, 
therefore it is difficult to realize a spin liquid state in real materials or realistic theoretical models. 
One possibility to realize a spin liquid state is to consider systems with strong frustrations. 
In fact this possibility is realized in the charge organic transfer salts $\kappa$-(BEDT-TTF)$_2\mathrm{X}$~\cite{shimizu03}. 
Another possibility is to consider systems which do not contain many electrons near the Fermi level. 
In this situation, the system might become insulating as the degree of the electron correlations increases, before it becomes an ordered state 
due to the cooperative correlations of the electrons near the Fermi level. 
  
\begin{figure}[t]
\begin{center}
\includegraphics[width=6.0cm,bb= 41 362 492 628]{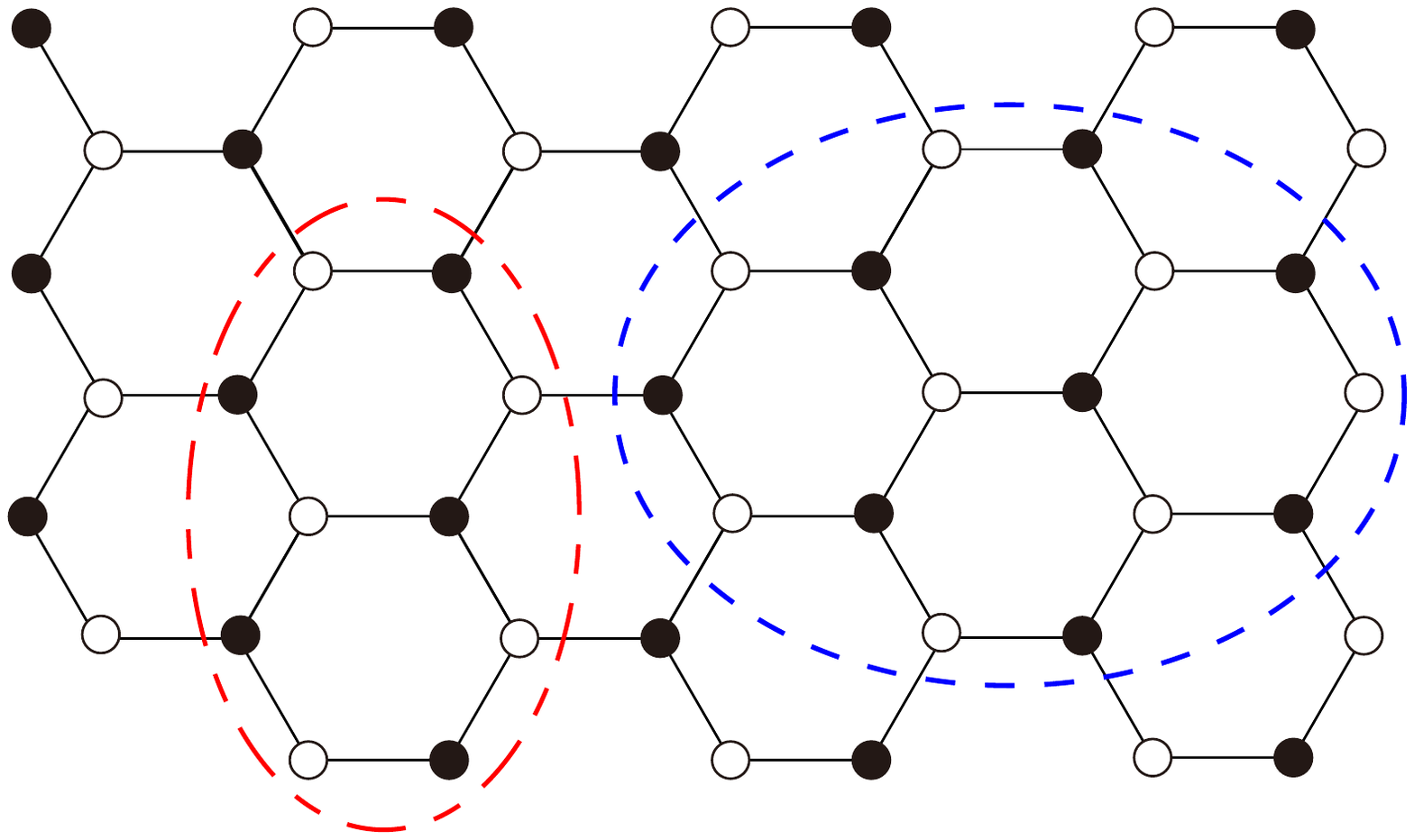}
\caption{
(Color online) 
The honeycomb lattice. 
%A pair of neighboring filled and unfilled sites is the unit cell. 
In the magnetic state the spins orient in the opposite directions 
on the filled and unfilled circles.  
The 10-site cluster circled by the dash-dotted line and 16-site cluster circled by the dotted line are used in our analysis. 
\label{fig:model}\\[-0.7cm]}
\end{center}
\end{figure}

\begin{figure}[t]
\begin{center}
\includegraphics[width=9.2cm,bb = 35 196 503 691]{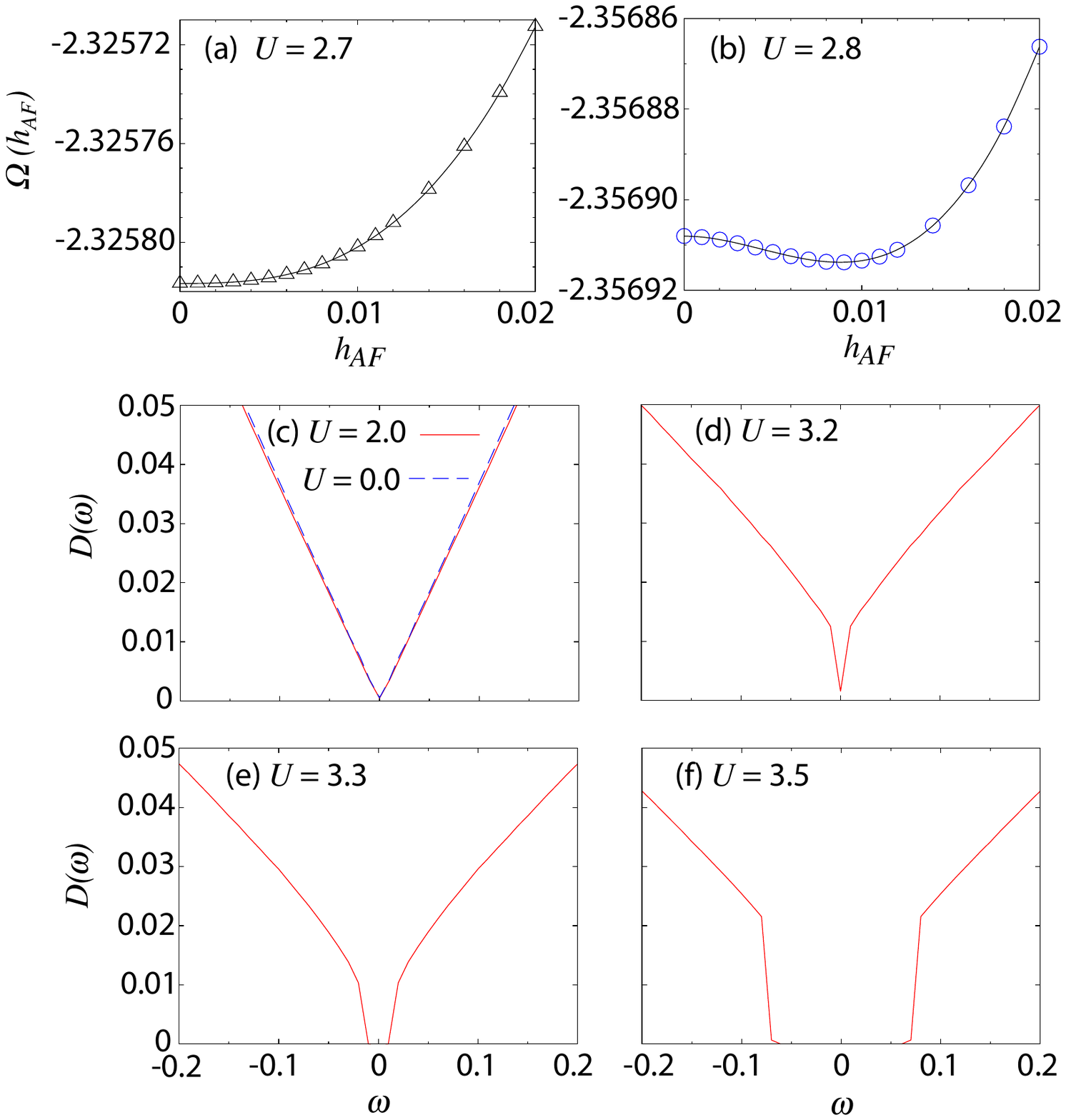}
\caption{
(Color online) (a)-(b) The grand potential per site $\Omega(h_{\rm AF})$ as a function of $h_{\rm AF}$ with $U=2.7$ (circles) and $U=2.8$ calculated on the 16-site-cluster. 
$\Omega(- h_{\rm AF}) = \Omega(h_{\rm AF})$ so only $ 0 \leq h_{\rm AF}$ is shown. 
(c)-(f) The density of states around the Fermi level $\omega=0$ computed on the 16-site cluster.   
\label{fig:omega-dos}\\[-0.7cm]}
\end{center}
\end{figure}

As an example of the latter possibility, the half-filled Hubbard model on the honeycomb lattice (See Fig.~\ref{fig:model}) 
has been extensively studied\cite{Meng,Sorella,assaad,toldin,yu,wu,seki,hassan,lieb,lau,ebato}. 
The Hamiltonian of this model is given by 
\begin{equation}
H = -t\sum_{\langle ij\rangle} c_{i\sigma }^\dag c_{j\sigma} + U \sum_{i} n_{i\uparrow} n_{i\downarrow} - \mu \sum_{i,\sigma} n_{i\sigma},
\label{eqn:hm}
\end{equation}
where $c^{(\dagger)}_{i\sigma}$ annihilates (creates) an electron with spin $\sigma$ on the site $i$, $n_{i\sigma}=c^{\dagger}_{i\sigma}c_{i\sigma}$
, $U$ is the on-site Coulomb repulsion, $\mu$ is the chemical potential, and $\langle ij\rangle$ denotes the nearest-neighbor pairs on the lattice.  
This is a simple model describing a graphene and there are also other materials exhibiting the hexagonal structures. 
When $U=0$ the band structure exhibits Dirac cones, and at half-filling the Fermi level crosses the conical vertex of the cones. 
So the density of state is zero only at the Fermi level and a semi-metal is realized. 
(See. Fig.~\ref{fig:omega-dos} (c).) 
%  for the density of states near the Fermi level at $U=0$.) 
Thus this model hosts the second possibility. 

So far the results of the preceding studies\cite{Meng,Sorella,assaad,toldin,yu,wu,seki,hassan,lieb,lau,ebato} seem to be rather controversial. 
Therefore we study this model again using the variational cluster approximation (VCA). VCA is classified as a kind of quantum cluster methods\cite{qcm}, and among the previous 
studies\cite{Meng,Sorella,assaad,toldin,yu,wu,seki,hassan,lieb,lau,ebato}, 
the quantum cluster methods, including VCA, are used in Refs. [\onlinecite{yu,wu,seki,hassan,lieb,lau,ebato}]. 
Therefore to clarify the significance of our study, we first briefly explain the quantum cluster methods and discuss the controversial situations up to now, mentioning the results of previous studies. 

In the quantum cluster methods, we divide the original infinite lattice into identical finite clusters and approximately compute physical quantities of the infinite system by solving a Hamiltonian defined on this finite size cluster. 
This Hamiltonian is called the cluster Hamiltonian. 
The cluster Hamiltonian is given by the same form as Eq.~(\ref{eqn:hm}) and now the indices $i,j$ run only $i,j=1,...,L_c$, where $L_c$ is the size of the finite cluster. 
When we investigate the spontaneous symmetry breaking we include appropriate Weiss field terms into this cluster Hamiltonian to resolve the degeneracy of the symmetry breaking directions. 
As an optional extension, $L_b$ sites are added at the boundary of the $L_c$-site cluster, 
which are called as ``bath'' sites and play a role to simulate the environments. 
When the bath sites are added, we set up the cluster Hamiltonian on the ($L_c + L_b$)-site cluster and include new hybridization hopping terms so that electrons can move 
between the original $L_c$-site cluster and bath sites. The electron correlations are neglected on these bath sites, therefore the on-site Coulomb interactions are absent on the bath sites in the cluster Hamiltonian.   
Then we approximate the electron correlations of the infinite system by the electron correlations among the $L_c$ correlated sites. 
To be precise, we approximate the self-energy of the infinite system by the exact self-energy of the $L_c$ sites of the cluster Hamiltonian defined on the 
($L_c, L_b$) cluster. 
The additional elements in the cluster Hamiltonian such as, parameters in the Weiss field terms and hybridization hopping between the correlated and bath sites, 
are determined based on self-consistency conditions or some variational principle. 
In principle the results of the above analysis converges to the exact results of the infinite system in the large $L_c$ limit with $L_b/L_c \rightarrow 0$. 
Here, the latter condition is necessary because there are no fictitious bath sites without electron correlations in the real system. 
In practice, the results depend on the cluster size and information on the infinite system is extracted using the cluster size dependence. 
In the present study, we need to compare the critical interaction strength $U_{\rm AF}$ of the magnetic transition and 
that of a non-magnetic metal-to-insulator transition $U_{\rm MI}$. 
In general, $U_{\rm AF}$ increases as the cluster size because the spatial fluctuations, which destabilize an ordered state, 
are taken into account more on a larger cluster. $U_{\rm MI}$ also increases as the cluster size. 
This is because the electron wave functions can spread wider region on a larger cluster, which lowers the kinetic energies and stabilizes a metallic state. 
Therefore we need to compute $U_{\rm AF}$ and $U_{\rm MI}$ at least on two clusters of different size and analyze their cluster size dependence.   

Next we briefly discuss the results of the previous investigations on this matter and explain the controversial situations. 
Some years ago, Meng {\textit{et al.}} \cite{Meng} found that $U_{\rm AF} = 4.3$ and $U_{\rm MI} = 3.5$ in model~(\ref{eqn:hm}) using the Quantum Monte Carlo simulations (QMC). 
Below $U\approx 3.5$, the system is a semi-metallic state, and beyond $U\approx 4.3$ it is an antiferromagnetic state,  
and in the range $3.5< U < 4.3$ a spin liquid state is realized. Their results inspired a lot of theoretical interests.  
Sorella {\textit{et al.}} \cite{Sorella} performed QMC for a system larger than Meng {\textit{et al.}} 
They found that $U_{\rm AF} = 3.8$ and the system is in a metallic state below $U_{\rm AF}$. Thus their result ruled out the 
existence of a spin liquid state. The two subsequent QMC studies\cite{assaad,toldin} support the result of Sorella {\textit{et al.}} 

The quantum cluster methods are also applied to this model. 
Yu {\textit{et al.}}\cite{yu} used VCA with the six-site hexagonal cluster and reported that a spin liquid is realized for $3.0 \lesssim U \lesssim 4.0$.   
Wu {\textit{et al.}}\cite{wu} used the cluster dynamical mean field theory (CDMFT) with clusters up to twenty-four sites and obtained e.g., $U_{\rm AF} = 3.7$ and $U_{\rm MI} = 3.2$ for the six-site hexagonal cluster, and 
$U_{\rm AF} = 3.8$ and $U_{\rm MI} = 3.2$ for a twenty-four site cluster. 
%These values show that the results converges fairly well.  
Both Yu {\textit{et al.}} and Wu {\textit{et al.}} argued that their results are in good agreement with Meng {\textit{et al.}} 
Seki and Ohta\cite{seki} studied this model using VCA and CDMFT and found that an insulating gap opens for infinitesimally small $U$ on some clusters while $U_{\rm AF}$ remains finite. So their results also support the existence of a spin liquid state. 
However, this gap which opens at infinitesimally small $U$ was further studied in Refs. [\onlinecite{hassan,lieb,lau,ebato}]  
and the accumulated results of these studies indicate that this gap does not correspond to the gap due to the electron correlations, 
but it is an unphysical artifact arising because the translational invariance is partially violated in the quantum cluster methods. 
Even though the precise condition is not yet clarified under which this unphysical gap opens at infinitesimally small $U$,  
so far this problem is observed only for some clusters in bipartite lattices.  
This unphysical gap pushes electrons away from Fermi level and makes the formation of an ordered state difficult. 
For example, a nesting at Fermi level is lost.   
Therefore we can not study $U_{\rm AF,MI}$ on a cluster with this problem. 
We need to analyze the cluster size dependence towards the limit $L_c \rightarrow \infty$ with $L_b/L_c \rightarrow 0$ 
selecting the clusters which do not have this pathological problem.  

Hassan and S\'en\'echal\cite{hassan} found that this unphysical gap is absent for the clusters  $(L_c, L_b) = (2,4)$ and $(4,6)$, 
and analyzed $U_{\rm AF,IM}$ 
using CDMFT and cluster dynamical impurity approximation(CDIA) on these clusters. 
They found that $U_{\rm AF} \simeq 3.3$ and $U_{\rm MI} \simeq 5.5$ for the $(L_c, L_b) = (2,4)$ cluster, and $U_{\rm AF} \simeq 4.0$ and $U_{\rm MI} \simeq 6.3$ for the $(L_c, L_b) = (4,6)$ cluster in CDMFT. 
In the case of CDIA, the metal-to-insulator transition becomes a first order transition and the coexisting range of the metallic and insulating phases are 
$5.7 \lesssim U \lesssim 6.0$ for  $(L_c, L_b) = (2,4)$, and $6.5 \lesssim U \lesssim 8.0$ for  $(L_c, L_b) = (4,6)$.  
Based on these results and observing that $U_{\rm MI}$ is much larger than $U_{\rm AF}$ they ruled out the existence of a spin liquid phase.  
However, their analysis always uses the clusters satisfying $L_b < L_c$ and largely violates the condition $L_c > L_b $.  
So it seems to be highly non-trivial if their results evaluate and approach towards the limit $L_c \rightarrow \infty$ with $L_b/L_c \rightarrow 0$. 
In fact, quantitatively, their $U_{\rm MI}$ on the $(L_c, L_b) = (4,6)$ cluster is 
already much larger than the results of Meng {\textit{et al.}} and Wu {\textit{et al.}}, and still increasing as the cluster size. 

In the case of VCA, $U_{\rm AF} \simeq 1.5$\cite{seki} and $U_{\rm MI} \simeq 3.4$\cite{lau} for the $(L_c, L_b) = (8,0)$ cluster, and 
$U_{\rm AF} \simeq 2.7$\cite{seki} and $U_{\rm MI} \simeq 3.0$\cite{lau} for the $(L_c, L_b) = (10,0)$ cluster. 
$U_{\rm MI}$ decreases rather largely as the cluster size increases, 
which is an anomalous behavior contradicting with the general argument of the cluster size dependence.  
Ebato {\textit{et al.}}\cite{ebato} found that a gap opens at infinitesimally small $U$ for $(L_c, L_b) = (12,0)$ cluster in VCA. 
Therefore proper information on the cluster size dependence of  $U_{\rm AF, MI}$ within VAC is still missing. 
We study the metal-to-insulator transition and magnetic transition by VCA using 
$(L_c, L_b) = (10,0)$ and $(16,0)$ clusters in Fig.~\ref{fig:model}. 
We found that $U_{\rm AF} = 2.7$ and $U_{\rm MI} = 3.0$ for the 10-site cluster, 
and $U_{\rm AF} = 2.7$ and $U_{\rm MI} = 3.2$ for the 16-site cluster. 
These results follow the general arguments of the cluster size dependence of the $U_{\rm AF,MI}$ and 
$U_{\rm MI} - U_{\rm AF}$ increases slightly as the cluster size. 
This result rules out the existence of a spin liquid state.  

{\it The formulation.}--- 
In VCA we use the thermodynamic grand potential 
$\Omega _{\mathbf{t}}$ written in the form of a functional of the self-energy $\Sigma $ as
\begin{equation}
\Omega _{\mathbf{t}}[\Sigma ]=F[\Sigma ]+\mathrm{Tr}\ln(-(G_0^{-1}-\Sigma )^{-1})
\label{eqn:omega}
\end{equation}%
for the Hamiltonian $H$ of the infinite system and for the sum of the cluster Hamiltonians defined on each cluster tiling the infinite lattice, 
which we denote as $H'$. In Eq. (\ref{eqn:omega}), $G_0$ is the non-interacting Green's function, $F[\Sigma ]$ is the Legendre 
transform of the Luttinger-Ward functional\cite{lw}, and the index $\mathbf{t}$ denotes the explicit dependence of 
$\Omega_{\mathbf{t}}$ on all the one-body operators in the Hamiltonian. 
This functional is stationary $\delta \Omega_{\mathbf{t}}[\Sigma ]/\delta \Sigma =0$ at the true self-energy of the corresponding Hamiltonian 
and this stationary condition is equivalent to the Dyson's equation. Bath sites are not used in VCA.  
All Hamiltonians with the same interaction part share the same functional form of $F[\Sigma ]$\cite{lw}. 
Since the interaction part, which is the on-site Coulomb repulsion, is the same for $H$ and $H'$, 
$F[\Sigma ]$ is the same for a given $\Sigma$ for $H$ and $H'$. 
So eliminating $F[\Sigma ]$, we obtain 
\begin{eqnarray}
\Omega_{\mathbf{t}}[\Sigma ]= \Omega' _{\mathbf{t'}}[\Sigma ] &+& \mathrm{Tr}\ln(-(G_0^{-1}-\Sigma )^{-1})
\nonumber 
\\ &-&\mathrm{Tr}\ln(-(G'_0{}^{-1}-\Sigma )^{-1}),
\label{eqn:omega3}
\end{eqnarray}%
where the prime denotes the quantities of $H'$. 
This is a functional relation between $\Omega _{\mathbf{t}}[\Sigma ]$ and $\Omega' _{\mathbf{t'}}[\Sigma ]$. 
By exactly solving $H'$, we compute the $\Omega _{\mathbf{t}}[\Sigma ]$ for the exact self-energy of $H'$. 
To investigate symmetry breaking, we include the Weiss terms into $H'$ and the parameters in the Weiss terms are determined by
solving the stationary condition. 
The exact self-energy of $H'$ satisfying the stationary condition, denoted as $\Sigma^{*}$, and 
$\Omega_{\mathbf{t}}[\Sigma^{*} ]$ are the approximate self-energy and grand-potential of $H$ in VCA. 
Physical quantities, such as expectation values of one-body operators, 
are evaluated using the Green's function $G_0{}^{-1}-\Sigma^{*} $. 
In VCA, the short-range correlations within the cluster are exactly taken into account and 
the restriction of the space of the self-energies $\Sigma$ into that of $H'$ is the only approximation. 

In the present case the Weiss field is given by 
%\begin{eqnarray}
%H_{\rm AF}&=& h_{\rm AF}\sum_{i} {\rm sign}(i)(n_{i\uparrow }-n_{i\downarrow }) 
$H_{\rm AF}= h_{\rm AF}\sum_{i} {\rm sign}(i)(n_{i\uparrow }-n_{i\downarrow })$ 
%\end{eqnarray}
with ${\rm sign}(i) = -1 (1)$ for the filled (unfilled) sites in Fig.~\ref{fig:model}. 
We set the chemical potential of the system and cluster as $\mu = \mu' = U/2$, which ensures the stationary half-filling solutions 
because the lattice is bipartite and particle-hole symmetry exists. 
The stationary condition is then reduced to $ d \Omega(h_{\rm AF})/ d h_{\rm AF} =0 $.  
In general, a non-magnetic solution with $ h_{\rm AF} =0$, and a magnetic solution with  $h_{\rm AF} \neq 0$ are obtained and 
we compare their energies to determine which is the stable ground state. 
In our study it is suffice to compare the grand potential per site $\Omega(h_{\rm AF})$, because $\mu$ is fixed to be $U/2$ at half-filling.  
To further investigate the properties of the ground state, we also compute 
the density of state per site 
\begin{eqnarray}
D(\omega)= \lim_{\eta \rightarrow 0}  \int%_{BZ}
{\frac{%
d^2 k }{(2\pi)^2 }}\frac{1}{n_c}\sum_{\sigma, a=1}^{n_c}\{ -\frac{1}{\pi} \mathrm{Im}G_{a\sigma}(k, \omega+i\eta) \},
\label{eqn:dos}
\end{eqnarray}
and the magnetic order parameter per site 
\begin{eqnarray}
M&=& \frac{1}{n_c}\sum_{a=1}^{n_c} \sum_{i} {\rm sign}(a)(n_{a\uparrow }-n_{a\downarrow }),
\end{eqnarray}
where $n_c = 2$ is the number of the sites in the unit cell in the sense of the sub-lattice formalism, 
and the $k$ integration is over the corresponding Brillouin zone. 
In Eq. (\ref{eqn:dos}), $\eta \rightarrow 0$ limit is evaluated using the standard extrapolation method 
by calculating $D(\omega)$ for $\eta =0.01$, $\eta/2$, and $\eta/4$. 

\begin{figure}[t]
\begin{center}
\includegraphics[width=8cm,bb = 147 330 454 511]{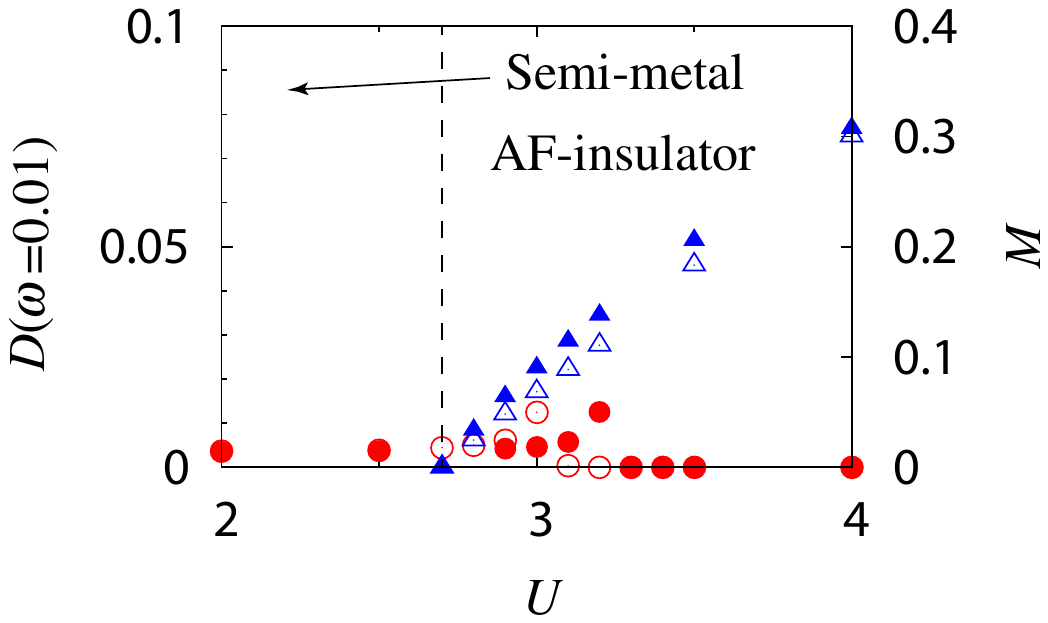}
\caption{
(Color online) The phase diagram computed on the 16-site cluster.
The filled circles are the density of state $D(\omega=0.01)$ and filled triangles are the magnetic order parameter $M$.
The unfilled marks correspond to the data computed on the 10-site cluster. 
\label{fig:phase}\\[-0.7cm]}
\end{center}
\end{figure}

{\it Results.}--- Fig.~\ref{fig:omega-dos} (a)-(b) show the grand potential per site $\Omega(h_{\rm AF})$ as a function of $h_{\rm AF}$ at 
(a) $U = 2.7$ and (b) $U = 2.8$ calculated on the 16-site cluster. 
A magnetic solution with $h_{\rm AF} \ne 0$ appears at $U = 2.8$ and the energy of this magnetic solution is lower than the non-magnetic 
solution with $h_{\rm AF}= 0$, so the magnetic state is realized at $U = 2.8$. 
For $U \leq 2.7$, a magnetic solution is not obtained and a non-magnetic state is realized. 
Fig.~\ref{fig:omega-dos} (c)-(f) show the density of state $D(\omega)$ around the Fermi level $\omega =0$ as a function of $\omega$ 
calculated on the 16-site cluster imposing $h_{\rm AF} =0$. 
For $U \leq 3.2$ the density of state is very close to zero only at $\omega =0$ around Fermi level and a semi-metallic nature is observed. 
For $U \geq 3.3$ a gap opens at Fermi level and the system is an insulator. 
By the similar analysis we found that the magnetic states are insulators. 
Fig.~\ref{fig:phase} shows the phase diagram computed on the 16-site cluster. 
The filled circles are the density of state $D(\omega=0.01)$ and filled triangles are the magnetic order parameter $M$.
$D(\omega=0)$ vanishes both in a semi-metal and insulator so $D(\omega=0.01)$ is shown to observe an insulating gap.  
The unfilled marks correspond to the data computed on the 10-site cluster. 
$U_{\rm AF} = 2.7$ and $U_{\rm MI} = 3.2$ on the 16-site cluster and $U_{\rm AF} = 2.7$ and $U_{\rm MI} = 3.0$ on the 10-site cluster. 
Our results of the 10-site cluster agree with the previous studies\cite{seki,lau,ebato}. 
Our results follow the general arguments on the cluster size dependence and in this sense anomalous behavior is not observed.  
The values of the order parameter suggests that the true minimum of the effective potential is slightly better simulated on the 16-site cluster. 
Both the magnetic and non-magnetic metal-to-insulator transitions are of the second order since there are no coexisting regions. 
The magnetic order parameter is $M \simeq 1.4$ around $U \simeq U_{\rm MI}$. 

{\it Summary and discussion.}---
In conclusion, we have studied the possible emergence of a spin liquid state in the Hubbard model on the honeycomb lattice by VCA. 
We computed the critical interaction strength $U_{\rm AF}$ of a magnetic transition and that of the non-magnetic metal-to-insulator transition $U_{\rm MI}$, and found that $U_{\rm AF} = 2.7$ and $U_{\rm MI} = 3.2$ on the 16-site cluster, and  
$U_{\rm AF} = 2.7$ and $U_{\rm MI} = 3.0 $ on the 10-site cluster.  
These results follow the general arguments of the cluster size dependence of the $U_{\rm AF,MI}$ and 
$U_{\rm MI} - U_{\rm AF}$ slightly increases as the cluster size. 
Therefore in this system, $U_{\rm MI}$ and $U_{\rm AF}$ are rather close and still $U_{\rm AF} < U_{\rm MI} $. 
The magnetic order parameter is not large but non-negligible in the region $ U \simeq U_{\rm MI}$. 
Thus a spin liquid state or a state very close to that is absent in this system. 
The analysis of the spectral weight function shows that the semi-metallic nature due to the Dirac cones are maintained up to the 
magnetic transition.

% If you have acknowledgments, this puts in the proper section head.
\begin{acknowledgments}
The author would like to thank H.~Fukazawa, H.~Kurasawa, H.~Nakada, T.~Ohama, Y.~Ohta, and K.~Seki for discussions. 
Parts of numerical calculations were done using the computer facilities of the IMIT at Chiba University, ISSP, and Yukawa Institute. 

\end{acknowledgments}

% Create the reference section using BibTeX:

\end{document}